%% file: main.tex
\documentclass[runningheads]{llncs}
\usepackage[T1]{fontenc}
\usepackage{graphicx}
\usepackage[breaklinks,colorlinks]{hyperref}
\usepackage{float}
\usepackage{booktabs}
\usepackage{multirow}
\usepackage{comment}

\usepackage{siunitx}

\usepackage{color}

\begin{document}
\title{All Sizes Matter: Improving Volumetric Brain Segmentation on Small Lesions}
\titlerunning{Improving Volumetric Brain Segmentation on Small Lesions}
%
\author{Ayhan Can Erdur\inst{1,2,*} \and
Daniel Scholz\inst{2,3,*} \and
Josef A. Buchner\inst{1,*} \and
Stephanie E. Combs\inst{1,4,5} \and
Daniel Rueckert\inst{2} \and
Jan C. Peeken\inst{1,4,5}}
\authorrunning{A. C. Erdur et al.}
%
\institute{Department of Radiation Oncology, Klinikum rechts der Isar, Technical University of Munich, Munich, Germany \and
Institute for Artificial Intelligence and Informatics in Medicine, Klinikum rechts der Isar, Technical University of Munich, Munich, Germany \and
Department of Neuroradiology, Klinikum rechts der Isar, Technical University of Munich, Munich, Germany \and
Deutsches Konsortium für Translationale Krebsforschung (DKTK), Partner Site Munich, Munich, Germany \and
Institute of Radiation Medicine (IRM), Department of Radiation Sciences (DRS), Helmholtz Center Munich, Munich, Germany \\
*contributed equally}
%
\maketitle              
\input{00_abstract}

\input{01_intro}
\input{02_methods}
\input{03_results}
\input{04_discussion}


%
%
%
\bibliographystyle{splncs04}
\bibliography{references}

\end{document}

%% file: 00_abstract.tex
\begin{abstract}
Brain metastases (BMs) are the most frequently occurring brain tumors.
The treatment of patients having multiple BMs with stereotactic radiosurgery necessitates accurate localization of the metastases.
Neural networks can assist in this time-consuming and costly task that is typically performed by human experts.
Particularly challenging is the detection of small lesions since they are often underrepresented in existing approaches.
Yet, lesion detection is equally important for all sizes.
In this work, we develop an ensemble of neural networks explicitly focused on detecting and segmenting small BMs.
To accomplish this task, we trained several neural networks focusing on individual aspects of the BM segmentation problem:  
We use blob loss that specifically addresses the imbalance of lesion instances in terms of size and texture and is, therefore, not biased towards larger lesions.
In addition, a model using a subtraction sequence between the T1 and T1 contrast-enhanced sequence focuses on low-contrast lesions.
Furthermore, we train additional models only on small lesions.
Our experiments demonstrate the utility of the additional blob loss and the subtraction sequence.
However, including the specialized small lesion models in the ensemble deteriorates segmentation results.
We also find domain-knowledge-inspired postprocessing steps to drastically increase our performance in most experiments.
Our approach enables us to submit a competitive challenge entry to the ASNR-MICCAI BraTS Brain Metastasis Challenge 2023.

\keywords{Metastasis segmentation \and Small lesion detection \and Brain MRI  \and Blob loss}
\end{abstract}

%% file: 01_intro.tex
\section{Introduction}

Brain metastases (BMs) occur approximately ten times more frequently than primary malignant brain tumors~\cite{Ostrom2018}.
In addition, nearly 10\% of patients with malignant tumors in the United States are expected to develop brain metastases \cite{Eichler2011}. 
While surgical resection is recommended for particularly large BMs, stereotactic radiosurgery (STS) is a possible treatment for patients with multiple BMs~\cite{Vogelbaum2022}. 
To minimize radiation exposure to healthy tissue, precise delineation of the target lesion in magnetic resonance imaging (MRI) is required for STS. 
Brain lesion delineation is a time-consuming task in clinical practice and also prone to interrater variability, i.e., deviations between annotators, especially for small BMs \cite{Sandstrom2018,Growcott2020}. 

Various machine learning algorithms \cite{ronneberger2015u,myronenko20193d} opened the possibility of automatic segmentation of such lesions. 
The potential of such approaches for automatic segmentation of primary brain tumors has been shown in previous brain tumor segmentation (BraTS) Challenges~\cite{Menze2015,Baid2021}: 
The neural network-based segmentations are not only faster but also independent of the rater. 
Moreover, recent research suggests that experts consistently score automatically created segmentations by neural networks higher than human-curated reference labels \cite{Kofler2021Metrics}. 

While only a small fraction of glioma patients suffer from multicentric lesions~\cite{Lasocki2016}, nearly 50\% of patients with BMs are affected by multiple metastases \cite{Fabi2011,Delattre1988}. 
This has a direct impact on measuring segmentation performance as well as on ranking contributions to segmentation challenges:
To evaluate large and small lesions equally, segmentation performance must be measured per lesion rather than cumulatively for all lesions combined.

The goal of this work is to develop an algorithm based on neural networks for the segmentation of the non-enhancing tumor core, enhancing tumor, and surrounding non-enhancing Fluid Attenuated Inversion Recovery (FLAIR) hyperintensity of BMs as a contribution to the ASNR-MICCAI BraTS Brain Metastasis Challenge~\cite{Karargyris2023Fed,Moawad2023}. 
We aim to improve the small lesion segmentation performance by employing special data augmentations, loss functions, and domain-knowledge-based postprocessing. 

%% file: 02_methods.tex
\section{Methods}
This section outlines the three key components of our challenge submission.
First, we describe the datasets provided for the challenge and our custom preprocessing and augmentations used.
Second, we describe our model and training configuration that serves as the basis for our submission.
Finally, we outline the improvements that were incorporated into the baseline, during, and after model training, with a focus on prediction plausibility and small lesion detection.

\subsection{Data}

\subsubsection{Datasets}

The provided data in this project consists of two parts:
238 patients provided directly by the BraTS challenge organizers \cite{Moawad2023} as well as 488 additional patients from the listed external datasets included in the challenge \cite{Rudie2023,oermann2022nyu}.

In total, four MRI sequences are supplied per patient as the following: pre-contrast T1-weighted sequence (t1w), post-contrast T1-weighted sequence (t1c),  T2-weighted sequence (t2w), and T2-weighted FLAIR sequence.

A segmentation map is also provided along MRI sequences to be the reference in training.
These maps consist of three labels: enhancing tumor (ET), non-enhancing tumor core (NETC), and surrounding non-enhancing FLAIR hyperintensity (SNFH).


\subsubsection{Preprocessing}

\begin{figure}[!h]
    \centering
    \includegraphics[width=\textwidth]{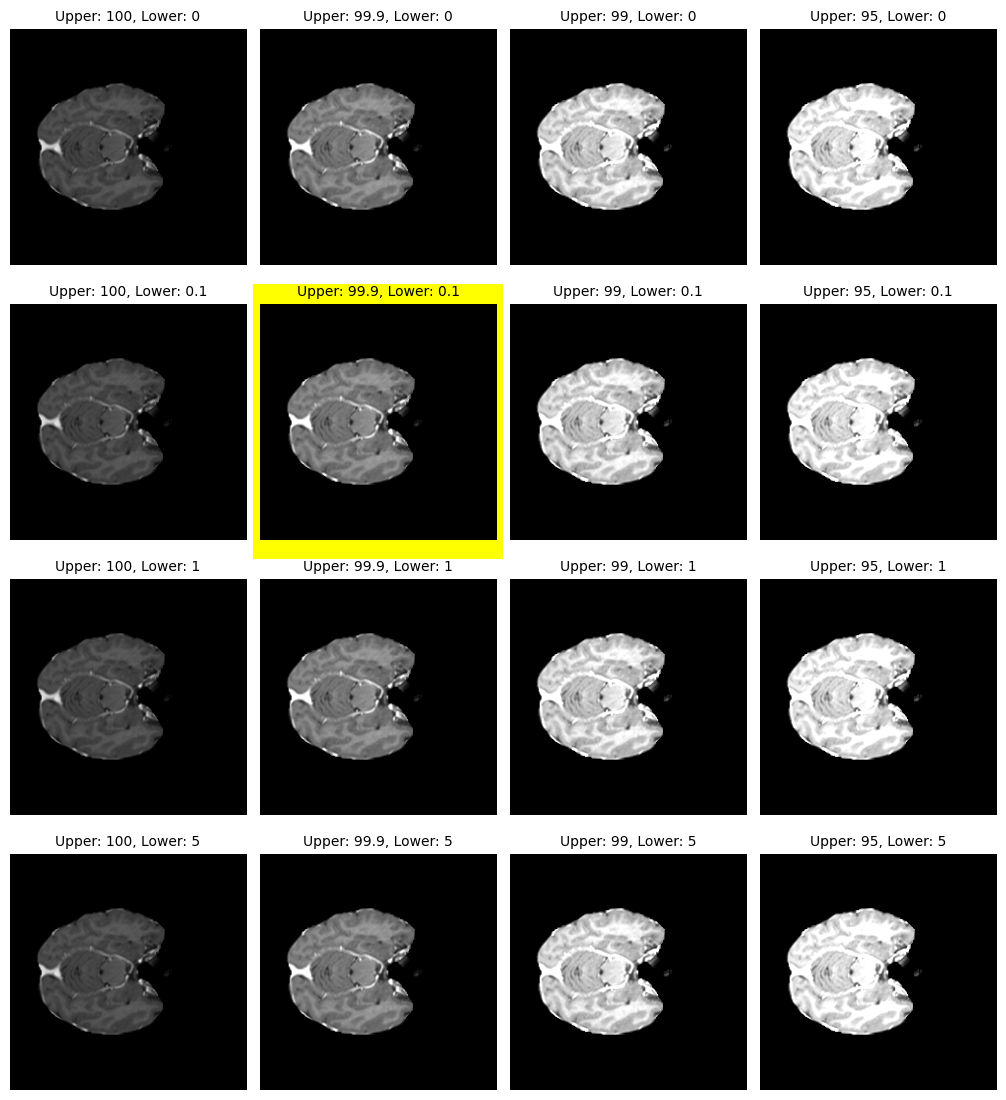}
    \caption{Comparison of different percentile-based intensity rescaling thresholds to obtain the best percentiles.
    We choose the 0.1th and the 99.9th as the lower and upper percentiles, respectively, (highlighted in yellow) to strike a balance between avoiding outliers and losing contrast.}
    \label{fig:intensity_scaling}
\end{figure}

All sequences were supplied as co-registered, skull-stripped sequences with an isotropic resolution of 1 millimeter in SRI24 space \cite{Rohlfing2010} with dimensions of 240 × 240 × 155 voxels.
We furthermore applied the following preprocessing steps to all samples:
We normalize the orientation to  (Left, Right), (Posterior, Anterior), (Inferior, Superior) and ensure 1mm isotropic resolution.
We scale intensities per channel to $\left[0,1\right]$ based on percentiles.
Best percentiles are determined visually by experts (\textit{cf.} Figure \ref{fig:intensity_scaling}).

To ensure gapless segmentations, we create 3-channel targets for our networks by merging the labels as whole tumor (WT), combination of ET, NETC, and SNFH, tumor core (TC), combination of ET and NETC, and enhancing tumor (ET).

\subsubsection{Data Augmentations}

Based on prior research in brain metastasis segmentation on a multi-center dataset of MRI images \cite{buchner2023development}, we determine a fixed set of data augmentations that are shared among all experiments.
The augmentations and the corresponding parameters are shown in Table \ref{tab:aug2}.

%
\begin{table}[h]
    \caption{Data augmentations and their corresponding parameters used during training.}
    \label{tab:aug2}
    \centering
\begin{tabular}{@{}lccc@{}}
\toprule
\textbf{Augmentation}                  & \textbf{Probability} & \multicolumn{2}{c}{\textbf{Parameters}}        \\ \midrule
\multirow{2}{*}{Random Flip}           & \multirow{2}{*}{0.5} & \multicolumn{2}{c}{\textbf{Axis}}              \\ \cmidrule(l){3-4}
                                       &                      & \multicolumn{2}{c}{0}                          \\ \midrule
\multirow{2}{*}{Random Affine}         & \multirow{2}{*}{0.5} & \multicolumn{2}{c}{\multirow{2}{*}{}}          \\
                                       &                      & \multicolumn{2}{c}{}                           \\ \midrule
\multirow{2}{*}{Random Gaussian Noise} & \multirow{2}{*}{0.5} & \textbf{Mean}      & \textbf{Std}              \\ \cmidrule(l){3-4}
                                       &                      & 0.0                & 0.1                       \\ \midrule
\multirow{2}{*}{Random Spatial Crops}  & \multirow{2}{*}{1}   & \textbf{Crop size} & $n_\mathrm{crops}$ \\ \cmidrule(l){3-4}
                                       &                      & [192,192,32]       & 2                         \\ \bottomrule
\end{tabular}
\end{table}
\subsection{Training}

\subsubsection{Base Model: SegResNetVAE}

\begin{figure}[h]
    \centering
    \includegraphics[width=\textwidth]{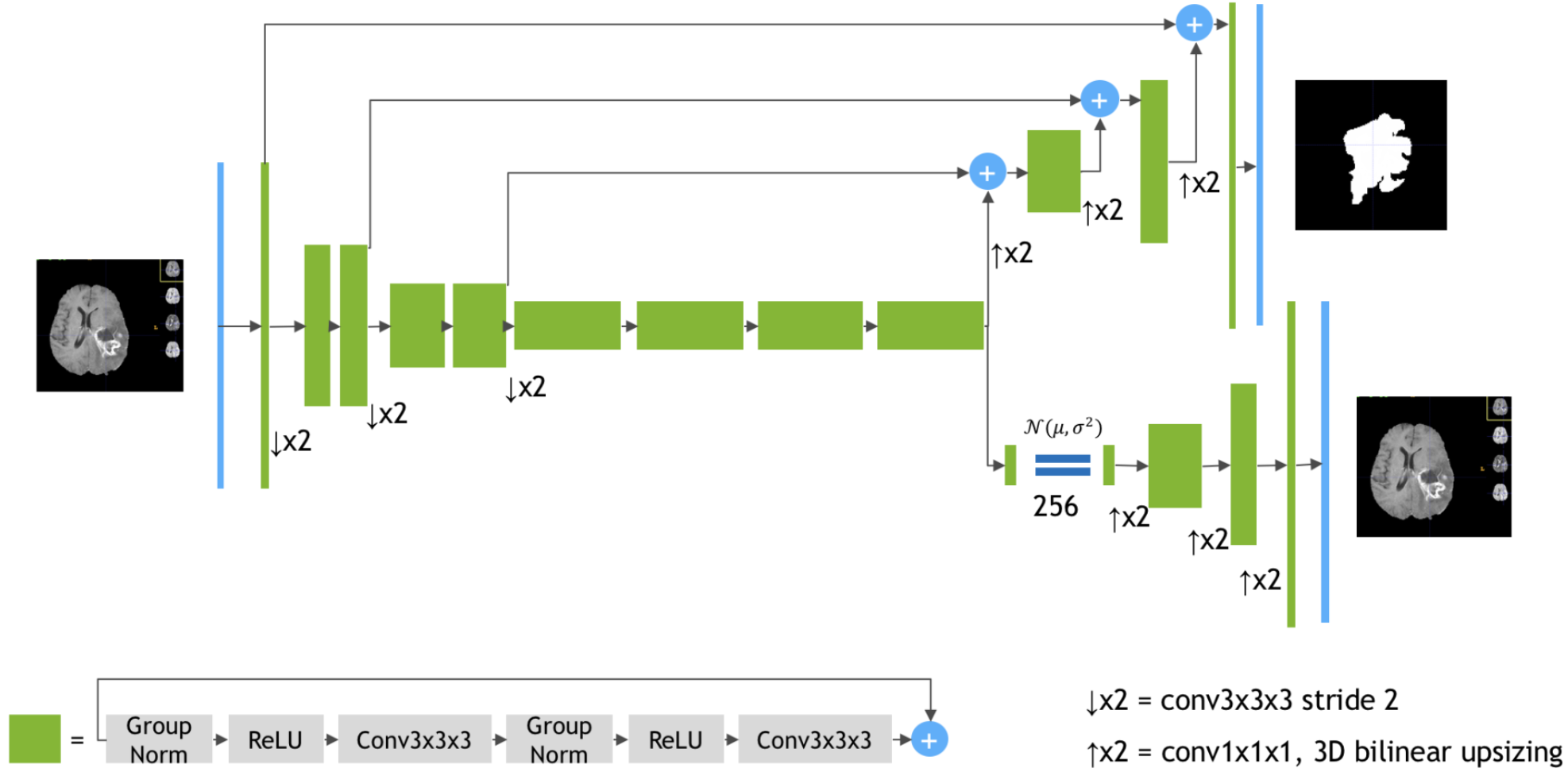}
    \caption{SegResNetVAE architecture with segmentation decoder at the top and variational autoencoder (VAE) branch for image reconstruction on the bottom. We replace the RELU activations in each residual block with Mish \cite{misra2019mish} functions. The input is a four-channel image of concatenated MRI sequences, and the output segmentation map has three channels representing WT, TC, and ET labels, respectively.}
    \label{fig:segresnet}
\end{figure}

For our experiments, we use the SegResNetVAE \cite{myronenko20193d} model, which is a 3D adaptation of the U-Net architecture \cite{ronneberger2015u} with modified residual blocks and a branch for image reconstruction using a variational autoencoder principle.
The additional branch functions as an auxiliary regularization on the learning task.
The model is visualized in Figure~\ref{fig:segresnet} with the number of blocks in each layer, the contents of residual blocks, and the upsampling and downsampling operations.

\subsubsection{Loss Functions}
In segmentation tasks, it is common practice to use the Dice loss~\cite{milletari2016v} or a weighted sum of the pixel-wise cross-entropy and the Dice loss (DiceCE) for training.
An equally weighted sum of the pixel-wise cross-entropy and the Dice loss is used as a loss function for our training runs.

\subsubsection{Training Configurations}
We first train a baseline model with default settings and incrementally add extra methods to the training or inference to improve the final segmentation performance.
Our baseline consists of the SegResNetVAE model with the DiceCE loss function.
For the input, all four of the available MRI sequences are concatenated to form a multi-channel image.
Table \ref{tab:train_config} describes the training hyperparameters that are shared among the baseline and the following models.

\begin{table}[h]
\caption{Hyperparameters used for training runs.}
\label{tab:train_config}
\centering
\begin{tabular}{@{}ccccc@{}}
\toprule
\multicolumn{3}{c}{\textbf{Optimizer}}            & \multirow{2}{*}{\textbf{Batch size}} & \multirow{2}{*}{\textbf{$n_\mathrm{epochs}$}} \\
Algorithm & Learning Rate & Weight Decay &                             &                                      \\ \midrule
AdamW     & \num{1e-4}    & \num{1e-5}   & 5                           & 200                                  \\ \bottomrule
\end{tabular}
\end{table}

As shown in Table~\ref{tab:aug2}, the images are cropped into smaller patches during training with two crops per patient, increasing the effective batch size from $5$ to $10$.
In the testing stage, we use the \textit{sliding window inference} method with an overlap of $0.75$ to obtain the segmentations matching the original input size.
We use MONAI \cite{Cardoso_MONAI_An_open-source_2022} library as the basis for the model implementations, loss functions, data preprocessing, and augmentation tools in our pipeline.

As a fixed setting for all experiments, we created a training, validation, and test split consisting of 80\%, 10\%, and 10\% portions, respectively, and used these splits for model tuning and comparison.

\subsection{Improvements}
In this section, we present steps taken to improve the segmentations overall and with special focus on small lesions compared to our baseline.

\subsubsection{Blob Loss}
Kofler \textit{et al.}~\cite{kofler2023blob} have shown that the Dice loss is biased toward larger lesions and performs poorly with an unbalanced set of instances, i.e., differences in size, texture, and morphology.
They have proposed a loss function, \textit{blob loss}, that addresses the imbalance by treating each lesion individually.
The blob loss functions as a wrapper around any segmentation loss to mask out all lesions but one, calculate the loss value per lesion, and compute the final loss by averaging the lesion losses of a patient.
With this method, they improve overall and instant-wise detection and segmentation performance in multi-lesion cases.

Following the suggestions of Kofler \textit{et al.}~\cite{kofler2023blob}, we always formulate the blob loss as an auxiliary term to DiceCE loss.
Throughout the paper, we refer to a weighted summation of the losses as the blob loss. This consists of a global DiceCE term and a lesion-focused blob loss term:

\begin{equation}
    \mathcal{L}_\mathrm{final} = 2 \cdot  \mathcal{L}_\mathrm{DiceCE} +  \mathcal{L}_\mathrm{blob}\left[\mathrm{Dice}\right]
\end{equation}
,with $\mathcal{L}_\mathrm{blob}\left[\mathrm{Dice}\right]$ depicting the vanilla Dice loss wrapped in blob loss as the lesion-wise evaluator.

To detect small lesions in BM patients and to achieve better instance-wise performance, we choose to use blob loss in our experiments. We also provide a comparison with DiceCE loss under the same settings.

\subsubsection{Subtraction Sequence}
Other works have included a subtraction sequence between t1w and t1c for BM segmentation~\cite{rudie2021three}, creating an image with highlighted contrast-enhancing lesions and almost zeroed out remaining tissue.
We adopt this domain knowledge and use the subtraction sequence $||\mathrm{t1c} - \mathrm{t1w}||^2$ as an additional channel in our input combination.

We keep the network architecture identical across experiments by swapping t2w with the subtraction sequence.
The t2w sequence contributes minimally to the results, such that identical or better performance can be achieved without it~\cite{Buchner2023Ablation}.

\subsubsection{Small Lesions Model}
Smaller lesions are generally overlooked by the original Dice loss formulation since larger errors in larger lesions contribute more to the score.
To overcome this shortcoming, we propose additional measures to improve the detection of small lesions.
By following a similar strategy to the protected group models approach by Puyol \textit{et al.}~\cite{puyol2021fairness}, we train a separate model for small lesions.
To this end, we mask out all lesions that are larger than a certain threshold $\tau$.
Since some of the samples end up with no remaining lesions, we filter these patients from the dataset for this specific training.
In our experiments, we set the lesion size thresholds to $\tau=1000$ voxels.

\subsubsection{Test Time Augmentations}
Buchner \textit{et al.}~\cite{buchner2023development} improve metastasis segmentation on MR images employing a set of test time augmentations.
Following their work, we apply additive Gaussian noise sampled from $\mathcal{N}\left(0,0.001\right)$ and flipping at random along the sagittal and coronal plane as test time augmentations.

\subsubsection{Ensembling}

Previous research shows that combining the predictions of multiple networks increases the segmentation performance and robustness~\cite{Kamnitsas2018ensemble}.
Therefore, we combine the predictions of the best-performing models into an ensemble for our final submission.
To achieve this, we compare a mean ensembling, i.e, averaging over the outputs, approach with the \textit{SIMPLE} \cite{Langerak2010Simple} algorithm implemented in the BraTS Toolkit \cite{kofler2020brats}, which is an iterative majority-voting approach.


\subsubsection{Postprocessing}
\begin{figure}
    \setlength{\abovecaptionskip}{1pt}
    \centering
    \includegraphics[width=0.8\textwidth]{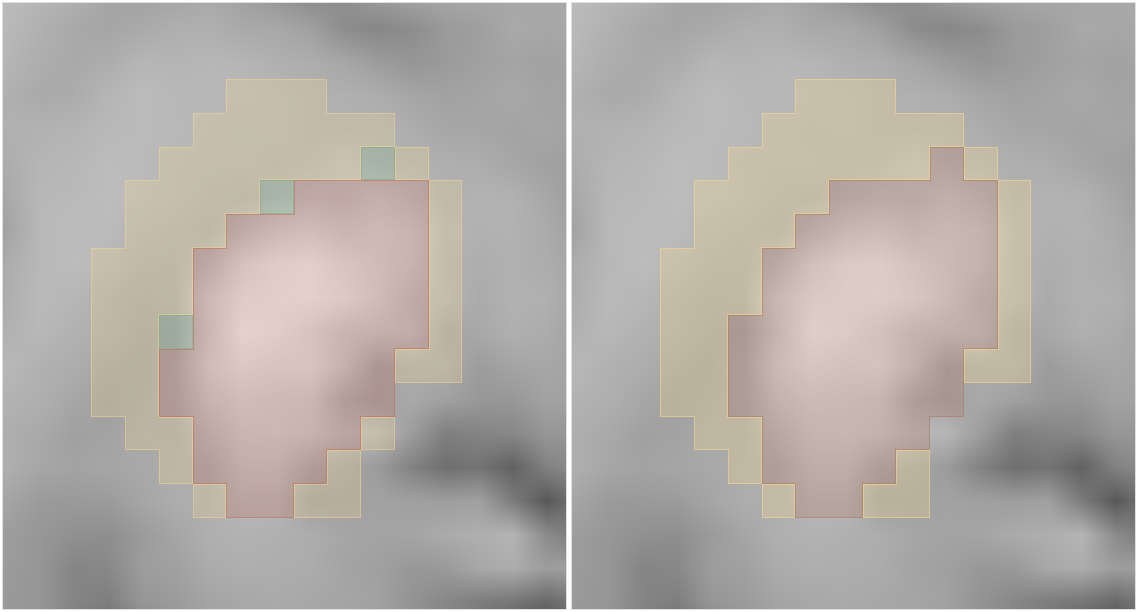}
    \caption{Comparison of an segmentation before (left) and after (right) postprocessing.
    Since the NETC (in green) is created by subtracting the ET (in red) from the TC, some border voxels may get labeled NETC.
    These get removed and labeled to ET by our postprocessing.}
    \label{fig:postprocessing}
\end{figure}

To aggregate the 3-channel outputs of our neural networks into a final segmentation map, we perform the inverse of input label merging, which is described in the preprocessing section.
Then, individual blobs of the respective labels are detected using a connected components analysis utilizing the Python library connected-components-3d \cite{cc3d2020github}.
To remove small lesions, which can be a result of the inversed label merging, we employ a postprocessing step based on the following rules:
\begin{itemize}
    \item WT: Blobs smaller than 25 voxels get removed.
    \item NETC: Blobs smaller than 20 voxels are added to the ET label.
    \item SNFH: Blobs smaller than 20 voxels get removed.
    \item ET: Blobs smaller than 10 voxels get removed.
\end{itemize}
An example of the resulting changes can be seen in Table~\ref{fig:postprocessing}.

\subsection{Metrics}
To determine the performance of our models, we use the evaluation script provided by the challenge organizers \cite{Moawad2023,valgithub}.
We report the cumulative Dice Similarity Coefficient (cDSC) across all lesions as well as the DSC per lesion (lDSC) for the ET, TC, and WT.
The 95th percentile of the Hausdorff Distance is also reported cumulatively (cHD95) and per lesion (lHD95).
For a more granular analysis of our experiments, we also measure false positives (FPs), i.e., additional detections, and false negatives (FNs), i.e., missed detections, averaged per patient.
Unless otherwise specified, we report the mean metric across all three labels.


%% file: 03_results.tex
\section{Results}

\subsection{Final Results}

We report our results in Table \ref{tab:final-results} on a hold-out test set from the provided challenge training data.
We find that both the additional blob loss and the subtraction sequence yield substantial improvement over the baseline.
Our contributions primarily cause a reduction in FP lesion detections, leading to better segmentations.
However, test time augmentations only reduce the number of FPs per patient at the expense of more FNs.
The DSC and the Hausdorff distance remain constant.
Combining the baseline model with the models with added blob loss and subtraction sequence using the iterative SIMPLE ensemble method, we improve the performance compared to the individual models and the mean ensemble.

\begin{table}
\centering
\caption{Results on our hold-out test set after postprocessing comparing our submission model to the baseline.} \label{tab:final-results}
    \centering
\begin{tabular}{l|cccccc}
\toprule
Model                  & lDSC ($\uparrow$)  & cDSC ($\uparrow$)  &  lHD95 ($\downarrow$)  & cHD95 ($\downarrow$)   &  FP ($\downarrow$)  &  FN ($\downarrow$)  \\ \midrule
Baseline & 0.43  & 0.69  &   267    &    32   &  0.75  &   0.63  \\
+ Blob Loss \cite{kofler2023blob}      & 0.46  & 0.70  &   \textbf{249}    &    \textbf{23}   &  0.69  &   0.63  \\
+ Sub. Seq. \cite{rudie2021three}  & 0.47  & 0.72  &   \textbf{249}    &    24   &  0.50 &  0.69 \\
+ TTA \cite{buchner2023development}              &  0.47  &  0.71  &   250    &    24   &    \textbf{0.41}  &  0.72 \\
\textbf{SIMPLE Ensemble \cite{Langerak2010Simple} } & \textbf{0.49}    &   \textbf{0.74}   &  254    &   \textbf{23}    &   0.57 &   \textbf{0.45}  \\ \bottomrule
\end{tabular}
\end{table}

\subsubsection{Preliminary Challenge Results}
We report our preliminary results on the challenge validation set obtained from the official challenge evaluation platform.
Our final model ensemble achieves a 0.47 average lDSC (0.45 ET, 0.50 TC, 0.47 WT) and a 158 average lHD95 (163 ET, 158 TC, 152 WT).





\subsection{Complementary Analysis of Our Proposed Methods}
We provide complementary analysis, justifying our choice of improvements leading to our main results.



\subsubsection{Postprocessing Based on Domain Knowledge}
By incorporating domain knowledge into our model predictions via postprocessing, we achieve better lesion-wise volumetric segmentation performance (\textit{cf.} Table \ref{tab:post_proc}). Furthermore, we are able to reduce the number of FPs at the expense of more FNs.

\begin{table}[h]
\centering
\caption{Comparing model performance with and without domain specific postprocessing (PP).
Our postprocessing improves primarily lesion-wise metrics compared to the unprocessed results.
We observe a drastic reduction in false positives indicated by the mean FP lesions per patient, while the FNs increase.
}\label{tab:post_proc}
\centering
\begin{tabular}{@{}l|cc|cc|cc|cc|cc|cc@{}}
\toprule
\multirow{2}{*}{Model} & \multicolumn{2}{c|}{lDSC ($\uparrow$)} & \multicolumn{2}{c|}{cDSC ($\uparrow$)} & \multicolumn{2}{c|}{lHD95 ($\downarrow$)} & \multicolumn{2}{c|}{cHD95 ($\downarrow$)} & \multicolumn{2}{c|}{FP ($\downarrow$)} & \multicolumn{2}{c}{FN ($\downarrow$)} \\
                                   &      & + PP &      & + PP &     & + PP &    & + PP &      & + PP  & & + PP \\ \midrule
Baseline & 0.38 & \textbf{0.43} & 0.69 & 0.69 & 286 & \textbf{267}  & \textbf{28} & 32   & 2.00 & \textbf{0.75} & \textbf{0.48} & 0.63 \\
+ Blob Loss \cite{kofler2023blob} & 0.35 & \textbf{0.46} & 0.70 & 0.70 & 281 & \textbf{249}  & 28 & \textbf{23} & 2.46 & \textbf{0.69} & \textbf{0.49} &  0.63\\
+ Sub. Seq. \cite{rudie2021three} & 0.38 & \textbf{0.47} & 0.72 & 0.72 & 270 & \textbf{249} & 27 & \textbf{24} & 1.94 & \textbf{0.50} & \textbf{0.52} & 0.69 \\
+ TTA  \cite{buchner2023development}     & 0.41 & \textbf{0.47} & \textbf{0.72} & 0.71 & 257	& \textbf{250} & 25	& \textbf{24} & 1.49 & \textbf{0.41} &  \textbf{0.55} & 0.72 \\
Mean Ensemble & 0.41 & \textbf{0.47} & \textbf{0.73} & 0.72 & 269 & \textbf{261} & \textbf{24} & 30 & 1.56 & \textbf{0.59} & \textbf{0.35} & 0.51 \\
SIMPLE \cite{Langerak2010Simple} & 0.43 & \textbf{0.49} & 0.74 & 0.74 & 261 & \textbf{254} & 23 & 23 & 1.40 & \textbf{0.57} & \textbf{0.34} & 0.45  \\
\bottomrule
\end{tabular}
\end{table}

\subsubsection{Ensembling Methods}
We evaluate different algorithms to combine the predictions of our best models into an ensemble.
Namely, we compare a mean ensemble to the \textit{SIMPLE} method implemented by BraTS Toolkit \cite{kofler2020brats}.
We show that the latter method performs better in both lesion wise and cumulative DSC (\textit{cf.} Table~\ref{tab:ensemble_comp}).

\begin{table}[h]
    \caption{Comparing the mean ensemble with the more sophisticated SIMPLE ensembling method~\cite{kofler2020brats}.}
    \label{tab:ensemble_comp}
    \centering
    \begin{tabular}{l|cccccc}
    \toprule
        Ensemble                  & lDSC ($\uparrow$)  & cDSC ($\uparrow$)  &  lHD95 ($\downarrow$)  & cHD95 ($\downarrow$)   &  FP ($\downarrow$)  &  FN ($\downarrow$)  \\ \midrule

    Mean       &  0.47  &  0.72  &   261    &    30     &    0.59   &  0.51 \\
    SIMPLE \cite{Langerak2010Simple} & \textbf{0.49}    &   \textbf{0.74}   &  \textbf{254}    &   \textbf{23}    &   \textbf{0.57} &   \textbf{0.45}  \\ \bottomrule
    \end{tabular}

\end{table}



\subsubsection{Small Lesion Networks}
We report segmentation results for the networks only trained on up to 1000 voxel-sized lesions on only these test images that contain lesions smaller than 1000 voxels (\textit{cf.} Table \ref{tab:small_lesions}).
The model performs worse with respect to general DSC and HD95 compared to the all-lesions-model.
To evaluate the impact of incorporating a small-lesion-model into our existing ensemble, we test the performance of an ensemble of our all-lesion-models models with the small-lesion-model.

In order to combine the small-lesion-model with the former ensemble and to obtain final predictions, a second ensembling step is introduced.
Namely, we apply the voxel-wise maximum operation between the outputs of the small-lesion-model and the all-lesion-models.
This ensemble did not improve the volumetric segmentation performance compared to the ensemble without the small lesion models while being computationally more expensive.
Consequently, we exclude the small lesion model from our final models for the challenge submission.

\begin{table}
\caption{Segmentation performance of models trained solely on small lesions. We report individual model performance and as an ensemble with our best models.}
\label{tab:small_lesions}
\centering
\begin{tabular}{@{}lc|cccccc@{}}
\toprule
\multicolumn{2}{l|}{Model} & lDSC ($\uparrow$) & cDSC ($\uparrow$) & lHD95 ($\downarrow$) & cHD95 ($\downarrow$) & FP ($\downarrow$) & FN ($\downarrow$) \\ \midrule
Small Lesions
                               & ($\tau=1000$) & 0.27 & 0.27 & 334 & 67 & 0.80 & 1.52 \\ \midrule
\multicolumn{2}{l|}{SIMPLE Ensemble \cite{Langerak2010Simple} }  & \textbf{0.49} & 0.74 & 254 & \textbf{23} & \textbf{0.57} & 0.45 \\
\multicolumn{2}{l|}{+ Small Lesions}  & 0.48 & 0.74 & \textbf{239} & 24 & 0.84 & \textbf{0.37} \\
\bottomrule
\end{tabular}
\end{table}

%% file: 04_discussion.tex
 \section{Discussion}
 
In this work, we developed an algorithm for automatically detecting and segmenting BMs. 
We added dedicated components to a baseline segmentation model focusing on the detection and segmentation of small lesions. 
Compared to a baseline model, we improved our lDSC by 14\%. 
Most of the improvement can be attributed to using blob loss, adding a subtraction sequence, and the model ensembling.

The introduction of blob loss improved both the lDSC and the cDSC, with the former being more affected.
This is likely due to a better segmentation quality of mainly small lesions, while the segmentation quality of larger lesions did not change significantly.

Using domain-specific postprocessing with fixed thresholds, we reduced the number of false positives by up to 75\%. 
For this work, we choose a threshold that balances specificity and sensitivity.
In contrast, in many clinical applications, the risk of missing BMs outweighs the additional manual effort of removing FPs, and the threshold should be adjusted accordingly. 
Ultimately, our competition entry aims to inspire the research community to improve the detection of small lesions in the future.

%% file: main.bbl
\begin{thebibliography}{10}
\providecommand{\url}[1]{\texttt{#1}}
\providecommand{\urlprefix}{URL }
\providecommand{\doi}[1]{https://doi.org/#1}

\bibitem{cc3d2020github}
Silversmith w, kemnitz n. 2020 seung-lab/connected-components-3d. seung-lab.
  see https://github.com/seung-lab/connected-components-3d.

\bibitem{Baid2021}
Baid, U., Ghodasara, S., Mohan, S., Bilello, M., Calabrese, E., Colak, E.,
  Farahani, K., Kalpathy-Cramer, J., Kitamura, F.C., Pati, S., Prevedello,
  L.M., Rudie, J.D., Sako, C., Shinohara, R.T., Bergquist, T., Chai, R., Eddy,
  J., Elliott, J., Reade, W., Schaffter, T., Yu, T., Zheng, J., Moawad, A.W.,
  Coelho, L.O., McDonnell, O., Miller, E., Moron, F.E., Oswood, M.C., Shih,
  R.Y., Siakallis, L., Bronstein, Y., Mason, J.R., Miller, A.F., Choudhary, G.,
  Agarwal, A., Besada, C.H., Derakhshan, J.J., Diogo, M.C., Do-Dai, D.D.,
  Farage, L., Go, J.L., Hadi, M., Hill, V.B., Iv, M., Joyner, D., Lincoln, C.,
  Lotan, E., Miyakoshi, A., Sanchez-Montano, M., Nath, J., Nguyen, X.V.,
  Nicolas-Jilwan, M., Jimenez, J.O., Ozturk, K., Petrovic, B.D., Shah, C.,
  Shah, L.M., Sharma, M., Simsek, O., Singh, A.K., Soman, S., Statsevych, V.,
  Weinberg, B.D., Young, R.J., Ikuta, I., Agarwal, A.K., Cambron, S.C.,
  Silbergleit, R., Dusoi, A., Postma, A.A., Letourneau-Guillon, L.,
  Perez-Carrillo, G.J.G., Saha, A., Soni, N., Zaharchuk, G., Zohrabian, V.M.,
  Chen, Y., Cekic, M.M., Rahman, A., Small, J.E., Sethi, V., Davatzikos, C.,
  Mongan, J., Hess, C., Cha, S., Villanueva-Meyer, J., Freymann, J.B., Kirby,
  J.S., Wiestler, B., Crivellaro, P., Colen, R.R., Kotrotsou, A., Marcus, D.,
  Milchenko, M., Nazeri, A., Fathallah-Shaykh, H., Wiest, R., Jakab, A., Weber,
  M.A., Mahajan, A., Menze, B., Flanders, A.E., Bakas, S.: The rsna-asnr-miccai
  brats 2021 benchmark on brain tumor segmentation and radiogenomic
  classification  (7 2021), \url{https://arxiv.org/abs/2107.02314v2}

\bibitem{buchner2023development}
Buchner, J.A., Kofler, F., Etzel, L., Mayinger, M., Christ, S.M., Brunner,
  T.B., Wittig, A., Menze, B., Zimmer, C., Meyer, B., et~al.: Development and
  external validation of an mri-based neural network for brain metastasis
  segmentation in the aurora multicenter study. Radiotherapy and Oncology
  \textbf{178},  109425 (2023)

\bibitem{Buchner2023Ablation}
Buchner, J.A., Peeken, J.C., Etzel, L., Ezhov, I., Mayinger, M., Christ, S.M.,
  Brunner, T.B., Wittig, A., Menze, B., Zimmer, C., Meyer, B., Guckenberger,
  M., Andratschke, N., Shafie, R.A.E., Debus, J., Rogers, S., Riesterer, O.,
  Schulze, K., Feldmann, H.J., Blanck, O., Zamboglou, C., Ferentinos, K.,
  Bilger, A., Grosu, A.L., Wolff, R., Kirschke, J.S., Eitz, K.A., Combs, S.E.,
  Bernhardt, D., Rückert, D., Piraud, M., Wiestler, B., Kofler, F.:
  Identifying core mri sequences for reliable automatic brain metastasis
  segmentation. medRxiv  \textbf{16},  2023.05.02.23289342 (7 2023).
  \doi{10.1101/2023.05.02.23289342},
  \url{https://www.medrxiv.org/content/10.1101/2023.05.02.23289342v2
  https://www.medrxiv.org/content/10.1101/2023.05.02.23289342v2.abstract}

\bibitem{Cardoso_MONAI_An_open-source_2022}
Cardoso, M.J., Li, W., Brown, R., Ma, N., Kerfoot, E., Wang, Y., Murray, B.,
  Myronenko, A., Zhao, C., Yang, D., Nath, V., He, Y., Xu, Z., Hatamizadeh, A.,
  Zhu, W., Liu, Y., Zheng, M., Tang, Y., Yang, I., Zephyr, M., Hashemian, B.,
  Alle, S., Zalbagi~Darestani, M., Budd, C., Modat, M., Vercauteren, T., Wang,
  G., Li, Y., Hu, Y., Fu, Y., Gorman, B., Johnson, H., Genereaux, B., Erdal,
  B.S., Gupta, V., Diaz-Pinto, A., Dourson, A., Maier-Hein, L., Jaeger, P.F.,
  Baumgartner, M., Kalpathy-Cramer, J., Flores, M., Kirby, J., Cooper, L.A.,
  Roth, H.R., Xu, D., Bericat, D., Floca, R., Zhou, S.K., Shuaib, H., Farahani,
  K., Maier-Hein, K.H., Aylward, S., Dogra, P., Ourselin, S., Feng, A.: {MONAI:
  An open-source framework for deep learning in healthcare}  (Nov 2022).
  \doi{https://doi.org/10.48550/arXiv.2211.02701}

\bibitem{valgithub}
Chung, V.: Brats 2023 lesion-wise performance metrics evaluation (2023),
  \url{https://github.com/rachitsaluja/brats_val_2023}

\bibitem{Delattre1988}
Delattre, J.Y., Krol, G., Thaler, H.T., Posner, J.B.: Distribution of brain
  metastases. Archives of Neurology  \textbf{45},  741--744 (7 1988).
  \doi{10.1001/ARCHNEUR.1988.00520310047016},
  \url{https://jamanetwork.com/journals/jamaneurology/fullarticle/587837}

\bibitem{Eichler2011}
Eichler, A.F., Chung, E., Kodack, D.P., Loeffler, J.S., Fukumura, D., Jain,
  R.K.: The biology of brain metastases—translation to new therapies. Nature
  Reviews Clinical Oncology 2011 8:6  \textbf{8},  344--356 (4 2011).
  \doi{10.1038/nrclinonc.2011.58},
  \url{https://www.nature.com/articles/nrclinonc.2011.58}

\bibitem{Fabi2011}
Fabi, A., Felici, A., Metro, G., Mirri, A., Bria, E., Telera, S., Moscetti, L.,
  Russillo, M., Lanzetta, G., Mansueto, G., Pace, A., Maschio, M., Vidiri, A.,
  Sperduti, I., Cognetti, F., Carapella, C.M.: Brain metastases from solid
  tumors: Disease outcome according to type of treatment and therapeutic
  resources of the treating center. Journal of Experimental and Clinical Cancer
  Research  \textbf{30}, ~1--7 (1 2011).
  \doi{10.1186/1756-9966-30-10/TABLES/5},
  \url{https://jeccr.biomedcentral.com/articles/10.1186/1756-9966-30-10}

\bibitem{Growcott2020}
Growcott, S., Dembrey, T., Patel, R., Eaton, D., Cameron, A.: Inter-observer
  variability in target volume delineations of benign and metastatic brain
  tumours for stereotactic radiosurgery: Results of a national quality
  assurance programme. Clinical oncology (Royal College of Radiologists (Great
  Britain))  \textbf{32},  13--25 (1 2020). \doi{10.1016/J.CLON.2019.06.015},
  \url{https://pubmed.ncbi.nlm.nih.gov/31301960/}

\bibitem{Kamnitsas2018ensemble}
Kamnitsas, K., Bai, W., Ferrante, E., McDonagh, S., Sinclair, M., Pawlowski,
  N., Rajchl, M., Lee, M., Kainz, B., Rueckert, D., Glocker, B.: Ensembles of
  multiple models and architectures for robust brain tumour segmentation.
  Lecture Notes in Computer Science (including subseries Lecture Notes in
  Artificial Intelligence and Lecture Notes in Bioinformatics)  \textbf{10670
  LNCS},  450--462 (2018). \doi{10.1007/978-3-319-75238-9_38/FIGURES/7},
  \url{https://link.springer.com/chapter/10.1007/978-3-319-75238-9_38}

\bibitem{Karargyris2023Fed}
Karargyris, A., Umeton, R., Sheller, M.J., Aristizabal, A., George, J., Wuest,
  A., Pati, S., Kassem, H., Zenk, M., Baid, U., Moorthy, P.N., Chowdhury, A.,
  Guo, J., Nalawade, S., Rosenthal, J., Kanter, D., Xenochristou, M., Beutel,
  D.J., Chung, V., Bergquist, T., Eddy, J., Abid, A., Tunstall, L., Sanseviero,
  O., Dimitriadis, D., Qian, Y., Xu, X., Liu, Y., Goh, R.S.M., Bala, S.,
  Bittorf, V., Puchala, S.R., Ricciuti, B., Samineni, S., Sengupta, E.,
  Chaudhari, A., Coleman, C., Desinghu, B., Diamos, G., Dutta, D., Feddema, D.,
  Fursin, G., Huang, X., Kashyap, S., Lane, N., Mallick, I., Mascagni, P.,
  Mehta, V., Moraes, C.F., Natarajan, V., Nikolov, N., Padoy, N., Pekhimenko,
  G., Reddi, V.J., Reina, G.A., Ribalta, P., Singh, A., Thiagarajan, J.J.,
  Albrecht, J., Wolf, T., Miller, G., Fu, H., Shah, P., Xu, D., Yadav, P.,
  Talby, D., Awad, M.M., Howard, J.P., Rosenthal, M., Marchionni, L., Loda, M.,
  Johnson, J.M., Bakas, S., Mattson, P.: Federated benchmarking of medical
  artificial intelligence with medperf. Nature Machine Intelligence 2023 5:7
  \textbf{5},  799--810 (7 2023). \doi{10.1038/s42256-023-00652-2},
  \url{https://www.nature.com/articles/s42256-023-00652-2}

\bibitem{kofler2020brats}
Kofler, F., Berger, C., Waldmannstetter, D., Lipkova, J., Ezhov, I., Tetteh,
  G., Kirschke, J., Zimmer, C., Wiestler, B., Menze, B.H.: Brats toolkit:
  translating brats brain tumor segmentation algorithms into clinical and
  scientific practice. Frontiers in neuroscience p.~125 (2020)

\bibitem{Kofler2021Metrics}
Kofler, F., Ezhov, I., Isensee, F., Balsiger, F., Berger, C., Koerner, M.,
  Demiray, B., Rackerseder, J., Paetzold, J.H., Li, H., Shit, S., Mckinley, R.,
  Piraud, M., Bakas, S., Zimmer, C., Navab, N., Kirschke, J., Wiestler, B.,
  Menze, B.: Are we using appropriate segmentation metrics? identifying
  correlates of human expert perception for cnn training beyond rolling the
  dice coefficient. Machine Learning for Biomedical Imaging  \textbf{2},
  27--71 (3 2021). \doi{10.59275/j.melba.2023-dg1f},
  \url{https://arxiv.org/abs/2103.06205v4}

\bibitem{kofler2023blob}
Kofler, F., Shit, S., Ezhov, I., Fidon, L., Horvath, I., Al-Maskari, R., Li,
  H.B., Bhatia, H., Loehr, T., Piraud, M., et~al.: blob loss: instance
  imbalance aware loss functions for semantic segmentation. In: International
  Conference on Information Processing in Medical Imaging. pp. 755--767.
  Springer (2023)

\bibitem{Langerak2010Simple}
Langerak, T.R., Heide, U.A.V.D., Kotte, A.N., Viergever, M.A., Vulpen, M.V.,
  Pluim, J.P.: Label fusion in atlas-based segmentation using a selective and
  iterative method for performance level estimation (simple). IEEE transactions
  on medical imaging  \textbf{29},  2000--2008 (12 2010).
  \doi{10.1109/TMI.2010.2057442},
  \url{https://pubmed.ncbi.nlm.nih.gov/20667809/}

\bibitem{Lasocki2016}
Lasocki, A., Gaillard, F., Tacey, M., Drummond, K., Stuckey, S.: Multifocal and
  multicentric glioblastoma: Improved characterisation with flair imaging and
  prognostic implications. Journal of clinical neuroscience : official journal
  of the Neurosurgical Society of Australasia  \textbf{31},  92--98 (9 2016).
  \doi{10.1016/J.JOCN.2016.02.022},
  \url{https://pubmed.ncbi.nlm.nih.gov/27343042/}

\bibitem{Menze2015}
Menze, B.H., Jakab, A., Bauer, S., Kalpathy-Cramer, J., Farahani, K., Kirby,
  J., Burren, Y., Porz, N., Slotboom, J., Wiest, R., Lanczi, L., Gerstner, E.,
  Weber, M.A., Arbel, T., Avants, B.B., Ayache, N., Buendia, P., Collins, D.L.,
  Cordier, N., Corso, J.J., Criminisi, A., Das, T., Delingette, H., Çağatay
  Demiralp, Durst, C.R., Dojat, M., Doyle, S., Festa, J., Forbes, F., Geremia,
  E., Glocker, B., Golland, P., Guo, X., Hamamci, A., Iftekharuddin, K.M.,
  Jena, R., John, N.M., Konukoglu, E., Lashkari, D., Mariz, J.A., Meier, R.,
  Pereira, S., Precup, D., Price, S.J., Raviv, T.R., Reza, S.M., Ryan, M.,
  Sarikaya, D., Schwartz, L., Shin, H.C., Shotton, J., Silva, C.A., Sousa, N.,
  Subbanna, N.K., Szekely, G., Taylor, T.J., Thomas, O.M., Tustison, N.J.,
  Unal, G., Vasseur, F., Wintermark, M., Ye, D.H., Zhao, L., Zhao, B., Zikic,
  D., Prastawa, M., Reyes, M., Leemput, K.V.: The multimodal brain tumor image
  segmentation benchmark (brats). IEEE Transactions on Medical Imaging
  \textbf{34},  1993--2024 (10 2015). \doi{10.1109/TMI.2014.2377694}

\bibitem{milletari2016v}
Milletari, F., Navab, N., Ahmadi, S.A.: V-net: Fully convolutional neural
  networks for volumetric medical image segmentation. In: 2016 fourth
  international conference on 3D vision (3DV). pp. 565--571. Ieee (2016)

\bibitem{misra2019mish}
Misra, D.: Mish: A self regularized non-monotonic activation function. arXiv
  preprint arXiv:1908.08681  (2019)

\bibitem{Moawad2023}
Moawad, A.W., Janas, A., Baid, U., Ramakrishnan, D., Jekel, L., Krantchev, K.,
  Moy, H., Saluja, R., Osenberg, K., Wilms, K., Kaur, M., Avesta, A., Pedersen,
  G.C., Maleki, N., Salimi, M., Merkaj, S., von Reppert, M., Tillmans, N.,
  Lost, J., Bousabarah, K., Holler, W., Lin, M., Westerhoff, M., Maresca, R.,
  Link, K.E., hoda Tahon, N., Marcus, D., Sotiras, A., LaMontagne, P.,
  Chakrabarty, S., Teytelboym, O., Youssef, A., Nada, A., Velichko, Y.S.,
  Gennaro, N., Students, C., of~Annotators, G., Cramer, J., Johnson, D.R.,
  Kwan, B.Y.M., Petrovic, B., Patro, S.N., Wu, L., So, T., Thompson, G., Kam,
  A., Perez-Carrillo, G.G., Lall, N., of~Approvers, G., Albrecht, J., Anazodo,
  U., Lingaru, M.G., Menze, B.H., Wiestler, B., Adewole, M., Anwar, S.M.,
  Labella, D., Li, H.B., Iglesias, J.E., Farahani, K., Eddy, J., Bergquist, T.,
  Chung, V., Shinohara, R.T., Dako, F., Wiggins, W., Reitman, Z., Wang, C.,
  Liu, X., Jiang, Z., Leemput, K.V., Piraud, M., Ezhov, I., Johanson, E.,
  Meier, Z., Familiar, A., Kazerooni, A.F., Kofler, F., Calabrese, E., Aneja,
  S., Chiang, V., Ikuta, I., Shafique, U., Memon, F., Conte, G.M., Bakas, S.,
  Rudie, J., Aboian, M.: The brain tumor segmentation (brats-mets) challenge
  2023: Brain metastasis segmentation on pre-treatment mri  (6 2023),
  \url{https://arxiv.org/abs/2306.00838v1}

\bibitem{myronenko20193d}
Myronenko, A.: 3d mri brain tumor segmentation using autoencoder
  regularization. In: Brainlesion: Glioma, Multiple Sclerosis, Stroke and
  Traumatic Brain Injuries: 4th International Workshop, BrainLes 2018, Held in
  Conjunction with MICCAI 2018, Granada, Spain, September 16, 2018, Revised
  Selected Papers, Part II 4. pp. 311--320. Springer (2019)

\bibitem{oermann2022nyu}
Oermann, M.L.E.: Nyumets\_brain v1.0 (3 2022),
  \url{https://nyumets.org/2022/03/29/nyumets-brain-v1/}

\bibitem{Ostrom2018}
Ostrom, Q.T., Wright, C.H., Barnholtz-Sloan, J.S.: Brain metastases:
  epidemiology. Handbook of Clinical Neurology  \textbf{149},  27--42 (1 2018).
  \doi{10.1016/B978-0-12-811161-1.00002-5}

\bibitem{puyol2021fairness}
Puyol-Ant{\'o}n, E., Ruijsink, B., Piechnik, S.K., Neubauer, S., Petersen,
  S.E., Razavi, R., King, A.P.: Fairness in cardiac mr image analysis: an
  investigation of bias due to data imbalance in deep learning based
  segmentation. In: Medical Image Computing and Computer Assisted
  Intervention--MICCAI 2021: 24th International Conference, Strasbourg, France,
  September 27--October 1, 2021, Proceedings, Part III 24. pp. 413--423.
  Springer (2021)

\bibitem{Rohlfing2010}
Rohlfing, T., Zahr, N.M., Sullivan, E.V., Pfefferbaum, A.: The sri24
  multichannel atlas of normal adult human brain structure. Human Brain Mapping
   \textbf{31},  798--819 (5 2010). \doi{10.1002/HBM.20906}

\bibitem{ronneberger2015u}
Ronneberger, O., Fischer, P., Brox, T.: U-net: Convolutional networks for
  biomedical image segmentation. In: Medical Image Computing and
  Computer-Assisted Intervention--MICCAI 2015: 18th International Conference,
  Munich, Germany, October 5-9, 2015, Proceedings, Part III 18. pp. 234--241.
  Springer (2015)

\bibitem{Rudie2023}
Rudie, J.D., MSE, R.S., MSE, D.A.W., Nedelec, P.M., Calabrese, E., Colby, J.B.,
  Laguna, B., Mongan, J., Braunstein, S., Hess, C.P., Rauschecker, A.M.,
  Sugrue, L.P., Villanueva-Meyer, J.E., Rudie, C.A.J.: The university of
  california san francisco, brain metastases stereotactic radiosurgery
  (ucsf-bmsr) mri dataset  (4 2023), \url{https://arxiv.org/abs/2304.07248v2}

\bibitem{rudie2021three}
Rudie, J.D., Weiss, D.A., Colby, J.B., Rauschecker, A.M., Laguna, B.,
  Braunstein, S., Sugrue, L.P., Hess, C.P., Villanueva-Meyer, J.E.:
  Three-dimensional u-net convolutional neural network for detection and
  segmentation of intracranial metastases. Radiology: Artificial Intelligence
  \textbf{3}(3),  e200204 (2021)

\bibitem{Sandstrom2018}
Sandström, H., Jokura, H., Chung, C., Toma-Dasu, I.: Multi-institutional study
  of the variability in target delineation for six targets commonly treated
  with radiosurgery. Acta Oncologica  \textbf{57},  1515--1520 (11 2018).
  \doi{10.1080/0284186X.2018.1473636/SUPPL_FILE/IONC_A_1473636_SM2616.ZIP},
  \url{https://www.tandfonline.com/doi/abs/10.1080/0284186X.2018.1473636}

\bibitem{Vogelbaum2022}
Vogelbaum, M.A., Brown, P.D., Messersmith, H., Brastianos, P.K., Burri, S.,
  Cahill, D., Dunn, I.F., Gaspar, L.E., Gatson, N.T.N., Gondi, V., Jordan,
  J.T., Lassman, A.B., Maues, J., Mohile, N., Redjal, N., Stevens, G., Sulman,
  E., van~den Bent, M., Wallace, H.J., Weinberg, J.S., Zadeh, G., Schiff, D.:
  Treatment for brain metastases: Asco-sno-astro guideline. Journal of Clinical
  Oncology  \textbf{40},  492--516 (2 2022). \doi{10.1200/JCO.21.02314}

\end{thebibliography}
